# A Practical Open-Source Software Stack for a Cloud-Based Quantum Computing System

Norihiro Kakuko*, Shun Gokita*, Naoyuki Masumoto[†], Keita Matsumoto[‡], Kosuke Miyaji[†], Takafumi Miyanaga[†], Toshio Mori[†], Haruki Nakayama[‡], Keita Sasada[§], Yasuhito Takamiya[§], Satoyuki Tsukano[†], Ryo Uchida[‡], and Masaomi Yamaguchi*

*Quantum Laboratory, Fujitsu LTD., Kanagawa, Japan
{kakuko.norihiro, gokita.shun, y.masaomi}@fujitsu.com

[†]Center for Quantum Information and Quantum Biology, Osaka University, Osaka, Japan
{masumoto.naoyuki, miyaji.kosuke, miyanaga.takafumi, t.mori, tsukano.satoyuki}.qiqb@osaka-u.ac.jp

[‡]Software Engineering Dept.1, Systems Engineering Consultants Co.,LTD., Tokyo, Japan
{matsumoto.keita, nakayama.haruki, uchida}@sec.co.jp

[§]Strategic Technology Center, TIS Inc., Tokyo, Japan
{sasada.keita, takamiya.yasuhito}@tis.co.jp

*Abstract*—Since the late 2010s, quantum computers have become commercially available, and the number of services that users can run remotely via cloud servers is increasing. In Japan, several domestic superconducting quantum computing systems, including our own, began operation in 2023. However, the design of quantum computing systems, especially in the most critical areas near quantum computers, remains largely undisclosed, creating a significant barrier to entry into the quantum computing field. If this situation continues, progress toward standardization, which is essential for guiding quantum computer development, will stall, and it will be difficult to develop a practical quantum computing system that can perform calculations on a supercomputer scale.

To address this issue, we propose Open Quantum Toolchain for OPerators and USers (OQTOPUS), a full-stack quantum computing system developed from research with real quantum computers. OQTOPUS is one of the world's largest open-source software projects, covering operational software from cloud-based execution environment construction to system operation. Furthermore, to perform quantum computing effectively and efficiently, it implements key features, such as transpilers, multi-programming, and error mitigation, in an area as close as possible to a quantum computer, an area that system vendors rarely disclose. Finally, this study presents experimental results of applying OQTOPUS to a real quantum computer. OQTOPUS is publicly available on GitHub and will notably lower the barrier to entry into the quantum computing field, contributing to the formation of a quantum computing developer community through open discussion.

## I. Introduction

Quantum computers have attracted attention from academia and industry for many years owing to their ability to perform calculations that are beyond the reach of classical computers. In 2016, IBM began to offer a cloud service, granting researchers and developers access to a quantum computer for the first time in the world [1]. Since then, several companies and national institutions have launched similar services [2]-[6]. In Japan, RIKEN, Fujitsu, and Osaka University have taken the lead in releasing three 64-qubit superconducting quantum computers as cloud services [4]-[6]. The growing availability of such services has accelerated research on quantum computing [7]-[9].

Although the number of cloud-based quantum computing systems is increasing [1]-[6], [10]-[14], the implementation of open-source software, especially in the most important area close to a quantum computer, remains limited. Therefore, the barrier is high for research institutes and companies interested in entering the quantum computing field by releasing their own hardware or offering cloud services using third-party quantum computers. Furthermore, without open-source software, the development of practical cloud-based quantum computing systems capable of performing calculations on a supercomputer scale remains unclear, making their standardization challenging.

To promote the standardization of a cloud-based quantum computing system and formation of a quantum computing developer community, we release Open Quantum Toolchain for OPerators and USers (OQTOPUS). OQTOPUS is one of the largest open-source software worldwide, offering one-stop access to quantum systems. We show the design of key features including transpilers, multi-programming, and error mitigation implemented in the rarely disclosed area close to a quantum computer, which can obtain states of qubits just before a quantum computation is executed, for performing computation effectively and efficiently. This paper first introduces the key features of OQTOPUS and then details the system design. Furthermore, by making our source codes as open as possible and sharing knowledge obtained during the development and operation, we facilitate open discussion toward building a practical quantum computing system. Our main contributions can be summarized as follows.

- We develop and release a full-stack quantum computing system. Using this system, we have begun operating a real quantum computer.

- We implement key features for quantum computing in the area that is close to a quantum computer and less publicly available.

Section II elaborates on these key features and highlights differences between conventional systems and OQTOPUS. Section III describes our main objective of making the full-stack

TABLE I. COMPARISON OF KEY FEATURES BETWEEN CONVENTIONAL SYSTEMS AND OQTOPUS.

| Systems / Key features | IBM Quantum | Amazon Braket | Azure Quantum | Qibo | qBraid | **OQTOPUS** |
|---|---|---|---|---|---|---|
| Server-side transpiler | ✓ (P)[a] | ✓ (N) | ✓ (N) | ✓ (P) | ✗ | **✓ (P)** |
| Server-side execution[b] | ✓ (N) | ✓ (N) | ✓ (N) | ✗ | ✗ | **✓ (P)** |
| Multi-programming[b] | ✗ | ✗ | ✗ | ✗ | ✗ | **✓ (P)** |
| Error Mitigation | ✓ (P) | ✓ (N) | ✓ (N) | ✓ (P) | ✓ (N) | **✓ (P)** |
| Estimation | ✓ (P) | ✓ (N) | ✗ | ✓ (P) | ✗ | **✓ (P)** |
| Composer[b] | ✓ (N) | ✗ | ✗ | ✗ | ✓ (N) | **✓ (P)** |

[a.] (P) and (N) mean public and non-public, respectively.
[b.] Although these features already exist, they are not publicly available in the popular conventional systems.

quantum computing system open. Section IV presents use cases enabled by OQTOPUS. Finally, Section V provides concluding remarks.

## II. KEY FEATURES OF A CLOUD-BASED QUANTUM COMPUTING SYSTEM FOR USERS

In this section, we describe the key features of a cloud-based quantum computing system that enable users to perform quantum calculations efficiently and effectively. The most fundamental feature in quantum computing is sampling, and conventional systems and OQTOPUS use similar features for receiving a user's request via cloud, processing it internally, and returning sampling results. Hence, this paper assumes that users desire feature enhancement that allows for the efficient and effective use of quantum computing, including sampling, the most. Although there are various features, we have prioritized them based on what will be more useful to users from perspective of being essential for many quantum algorithms, improving calculation accuracy of current quantum computers suffering from errors, and eliminating difficulty for users to use quantum computers. Specifically, we have started implementing the following six key features.

**Server-Side Transpiler:** A transpiler converts and optimizes quantum circuits created by users so that they can be executed on the target quantum computer. Because the computational costs become enormous as the size of the quantum circuits increases, it is desirable to run the transpiler on a service provider's server.

**Server-Side Execution:** Using this feature, the service provider's server runs an entire quantum–classical hybrid algorithm, iterating the sampling of a quantum circuit whose parameters are tuned with a classical computer based on the previous sampling result. Without this feature, users will have to wait in a queue on a quantum computer for every sampling request.

**Multi-Programming:** This feature runs multiple quantum circuits in parallel on a single quantum chip.

**Error Mitigation:** This feature reduces the negative effect of errors that occur in a quantum computer.

**Estimation:** Estimating an expectation value means to calculate $\langle\psi|H|\psi\rangle$ for a Hamiltonian $H$ and a quantum state $|\psi\rangle$ after the sampling of the target quantum circuit. It is useful in quantum efficiency and quantum approximate optimization algorithm, which performs a variational optimization based on the expectation value of the Hamiltonian.

**Composer:** This feature is a tool that allows users to design, visualize, and run quantum circuits. Users can easily create their quantum circuits with it.

Table I compares these six features between popular conventional systems and OQTOPUS. Table I confirms that our system has all features, while IBM, Amazon, Azure, and Qibo have more than half of the features, but not all. Although many features in these conventional systems are not publicly available (N), OQTOPUS makes all features publicly available (P), as shown in Table I. In our system, features other than the composer are implemented in the area close to a quantum computer for efficient computing and their increased usefulness to users.

## III. OPEN QUANTUM TOOLCHAIN FOR OPERATORS AND USERS (OQTOPUS)

Our main contributions are the development of a full-stack quantum computing system based on OQTOPUS, including the key features discussed in the previous section, and making it available to all users [15]. As shown in Fig. 1, OQTOPUS operates across three layers. In the frontend layer, where computation is performed on a user's computer, QURI Parts OQTOPUS sends quantum jobs submitted by a user to a cloud layer. In the cloud layer that receives these quantum jobs, OQTOPUS Cloud is responsible for job and user managements. A user can check the status of their quantum jobs and hardware information, such as qubit fidelity, via a web user interface (UI) of OQTOPUS Frontend. OQTOPUS Engine in the backend layer retrieves jobs managed in OQTOPUS Cloud and executes quantum programs, working with Tranqu Server, providing the transpilers and Device Gateway, which serves as an interface for connecting to a pulse controller. For system operation, Qdash performs calibration, which is essential for a quantum computer, and visualizes results; meanwhile, OQTOPUS Admin facilitates user management via a web UI for system operators.

As mentioned above, OQTOPUS provides all basic operational software required for a cloud-based quantum computing system. In all software, we tend to use open and standard technologies because their issues have often been identified and addressed. Furthermore, the tools and documentation of these technologies are often well developed, facilitating their easy introduction. Therefore, adopting open and standard technologies leads to efficient development of quantum computing systems, an increase in the number of users, and the securing of human resources. The details of our software are explained in the following subsections A–D.

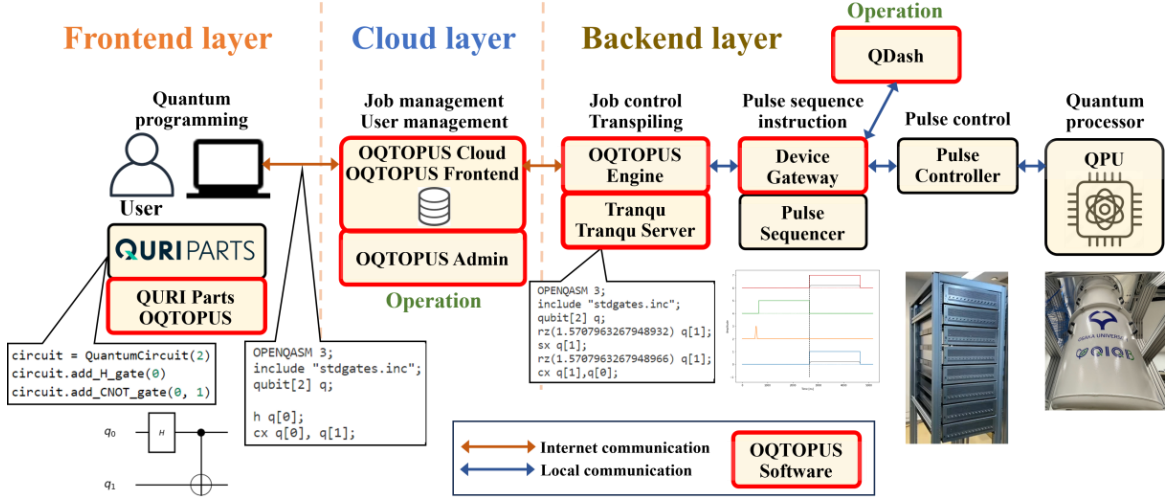

Fig. 1. Overview of OQTOPUS.

*A. Frontend Layer*

The frontend layer is where users directly perform quantum programming, so its ease of use is important. This is directly related to a low learning cost, a wide range of available libraries, and community support. Furthermore, quantum programming libraries are often installed and used by users in their own environments, while quantum computing system providers lack access to the users' environments, making it difficult to enforce library versions. To solve this problem, we provide the following QURI Parts OQTOPUS, including popular libraries with stable versions.

*1) QURI Parts OQTOPUS*: Our system adopts the quantum programming library QURI Parts [16], an open-source software written in Python [17]. Python is widely used to develop quantum programming libraries, including Qiskit [18], owing to its extensive scientific computing libraries, such as NumPy [19] and SciPy [20], and high compatibility with the pre and postprocessing of quantum computing. Therefore, it is reasonable that QURI Parts is written in Python. QURI Parts includes the best versions of popular libraries that provide a stable environment. A quantum circuit programmed with QURI Parts is converted to a code written in QpenQASM 3 [21], which is the most widely used intermediate representation for describing quantum circuits and is considered highly versatile. A quantum job containing the converted code is sent to OQTOPUS Cloud.

*B. Cloud Layer*

The cloud layer is built as an application stack on Amazon Web Services (AWS), a well-known public cloud service. Adopting a serverless architecture [22] allows our system to scale flexibly in response to varying workloads while reducing infrastructure operational costs. Although this layer is exposed to external networks and susceptible to cyberattacks, the cost associated with the cyberattack risk can be reduced by leveraging appropriate AWS services [23]. The cloud layer serving as a bridge between the frontend and backend layers exposes its public interface defined in OpenAPI Specification [24], enabling seamless communication between various frontend applications and quantum devices, including third-party ones.

*1) OQTOPUS Cloud*: OQTOPUS Cloud is a core component of the cloud layer, providing features, such as job and user managements. It receives quantum jobs only from authenticated users, pushes these jobs to a job queue for the backend layer to process, and manages the storage for quantum computations. The relationship between the cloud and backend layers is assumed to be one-to-many, handling multiple quantum computers. As mentioned earlier, OQTOPUS Cloud is a collection of cloud-based software: Amazon Cognito [25] handles user authentication and authorization, Amazon API Gateway [26] publishes Application Programming Interfaces (APIs), AWS Lambda [27] manages the execution and deployment of server-side applications, and Amazon Relational Database Service [28] stores quantum jobs.

*2) OQTOPUS Frontend*: The OQTOPUS Frontend is a web application that serves as a GUI for writing and managing quantum jobs. All users of our system must sign up through the OQTOPUS Frontend before using quantum computers. Additionally, authenticated users can issue API keys via the OQTOPUS Frontend. With a valid API key, users can interact with our system from other frontend services besides the OQTOPUS Frontend itself. As a programming environment, this application offers users two ways to write quantum programs: the quantum circuit composer and a text-based code editor. The former allows users to create their quantum programs by visually composing quantum circuits, while the latter accepts code conforming to the OpenQASM 3 specification. In a typical setting, the OQTOPUS Frontend is deployed on AWS Amplify [29], a platform for building and hosting web applications on AWS-managed infrastructure.

*C. Backend Layer*

The backend layer should be operated in facilities where a quantum computer is installed. Because features that depend on the quantum computer states, such as qubit connectivity and fidelity, are implemented in this layer, having it in a facility makes it easy to deal with a change in the states. As an execution platform for quantum computing, it works with the cloud layer, Tranqu Server, and Device Gateway to perform quantum programs as follows.

*1) OQTOPUS Engine:* OQTOPUS Engine retrieves quantum jobs from the cloud layer, schedules their execution at appropriate time, and communicates with Device Gateway to perform quantum computations. To handle quantum circuit transpilation, it uses Tranqu, a framework that runs various transpilers via Tranqu Server. Because current quantum computers allow only one job to run at a time on a quantum chip, our system follows a first-in-first-out queue and executes quantum jobs sequentially. The engine also handles various processes required for quantum computing, including key features such as server-side execution, multi-programming, error mitigation, and estimation. Each feature is compiled separately, forming a microservice architecture through protocols, such as gRPC [30].

*a) Server-Side Execution*: This feature allows users to execute Python programs that implement quantum–classical hybrid algorithms in the backend layer. Users submit programs to the cloud layer instead of quantum circuits. The submitted programs are then downloaded to OQTOPUS Engine and executed. While a program is running, it will not be requeued and other jobs will not interrupt it. Thus, it is possible to run the iteration of a quantum–classical hybrid algorithm, occupying the quantum chip. Although similar mechanisms for the efficient execution of a quantum–classical hybrid algorithm exist in other quantum computing systems, such as IBM Quantum and Amazon Braket, as shown in the previous section II, they provide priority execution and other jobs may interrupt the execution of a quantum–classical hybrid algorithm. Meanwhile, our server-side execution occupies the quantum chip and cannot be interrupted by other jobs.

*b) Multi-Programming*: This feature is a technique for performing quantum computation by combining multiple quantum circuits into a single circuit to efficiently use qubits [31]-[33]. Although multi-programming has been used to improve the accuracy of certain quantum algorithms [34], it is implemented in our system to improve the throughput of a quantum computer. Specifically, our multi-programming is executed when a user submits multiple quantum circuits to the cloud layer in the form of an array described in the definition that we have given. Then, OQTOPUS Engine combines those quantum circuits into a single quantum circuit and executes it on a quantum chip. Results obtained are then divided into individual outcomes for each original quantum circuit by OQTOPUS Engine. The current implementation is simple, as described above, where a user manually specifies multiple quantum circuits to be executed simultaneously, and OQTOPUS Engine combines and executes them. An automatic combining mode of multiple quantum circuits requested by different users is also planned for implementation.

*c) Error Mitigation*: Environmental noise and gate imperfections occur in a quantum chip, reducing the quality of quantum computing. Although the development of fault-tolerant quantum computers in the future may solve this issue, the error rate of current quantum computers is high for performing quantum computations. To address these challenges, our system employs a quantum error mitigation technique that extracts optimal signals from noisy measurement data. The technology implemented in our system is based on a Qiskit's 1-qubit tensor product readout error mitigator [18] and is packaged to run on the service provider's server. Unlike Qiskit alone, which is unstable owing to its dependency on the user's environment, our solution facilitates effective error mitigation that is automatically tuned to real-time error conditions observed on quantum computers.

*d) Estimation*: The expectation value $\langle \psi | H | \psi \rangle$ is calculated by decomposing the given Hamiltonian $H$ into a linear sum of the direct products of Pauli operators. The obtained expectation value for each Pauli operator is the sampling result of each qubit, and these values are summed on a classical computer. Qiskit's BackendEstimator [18] that enables expectation value estimation is packaged to provide as Estimation. Rather than simply using Qiskit, we have implemented this feature so that users can use error mitigation described above when performing sampling. Because the process is performed at the server side, users do not need to requeue for repeated sampling. The transpilers for quantum circuits are provided through Tranqu and Tranqu Server, enabling users to take the full advantage of the Tranqu's benefits, as shown in the next subsection.

*2) Tranqu*: Tranqu is a one-stop transpiler framework that allows users to leverage various quantum programming libraries and formats for quantum circuits. The transpilers for a quantum circuit are required to solve an NP-complete problem and typically depend on heuristic algorithms. Consequently, the quality of quantum circuits generated by each transpiler (such as the gate count, circuit depth, and execution efficiency) varies notably. The main advantage of Tranqu is that it enables users to process the target circuit using multiple transpilers and compare their results. Thus, users can select the most efficient circuit for execution on the target quantum computer. Tranqu simplifies the process of quantum circuit conversion, allowing users to focus instead on the quality of results. Currently, Qiskit's transpiler [18] and ouqu-tp's transpiler developed by Osaka University [35] are already implemented in Tranqu. Plans are underway to include other popular transpilers in the future.

*3) Tranqu Server*: Tranqu Server is a transpiler service application built with Tranqu. Transpilers consume more computational resources as the size of quantum circuits

```
Listing 1 Sampling Job
1  from quri_parts.circuit import QuantumCircuit
2  from quri_parts_oqtopus.backend import OqtopusSamplingBackend
3
4  circuit = QuantumCircuit(1)
5  circuit.add_H_gate(0)
6
7  backend = OqtopusSamplingBackend()
8  job = backend.sample(
9      circuit,
10     device_id="anemone",
11     name="sampling-001",
12     description="Experiment of sampling",
13     shots=1000
14 )
```

```
Listing 2 Estimation Job
1  from quri_parts.circuit import QuantumCircuit
2  from quri_parts_oqtopus.backend import OqtopusEstimationBackend
3  from quri_parts.core.operator import Operator, pauli_label
4
5  circuit = QuantumCircuit(2)
6  circuit.add_CNOT_gate(0, 1)
7
8  operator = Operator({
9      pauli_label("X 0 X 1"): 1.5,
10     pauli_label("Y 0 Z 1"): 1.2
11 })
12
13 backend = OqtopusEstimationBackend()
14 job = backend.estimate(
15     circuit,
16     device_id="anemone",
17     name="estimation-001",
18     description="Experiment of estimation",
19     operator=operator,
20     shots=1000
21 )
```

increases. Therefore, service providers may prefer to offer the transpilers as a cloud service, such as Qiskit Transpiler Service, rather than running them on the user's environment. Tranqu Server provides access to various transpilers via protocols, such as gRPC [30]. Using Tranqu Server, service providers can integrate Tranqu as a service in their systems and run the transpilers efficiently. Its current implementation supports gRPC servers, with plans to include RESTful API [36] support and additional features in the future.

*4) Device Gateway*: This server converts the OpenQASM 3 code of a quantum circuit into commands compatible with various quantum control software platforms for communication with quantum computers. To map virtual qubits specified in the OpenQASM 3 code to physical qubits when a quantum computer performs computation, this server also holds calibration data and device topology information. gRPC is used for efficient remote procedure calls owing to its modular backend architecture, which supports seamless integration with various quantum control software and simulators. Currently, Qulacs [37] is used as the simulator backend. Furthermore, this server works closely with OQTOPUS Engine to perform quantum jobs.

*D. Operation*

In large-scale software development like ours, if the development and operation teams do not cooperate, it is common to see features being developed without considering operational environment or delays in resolving operational issues occurring in the operational environment. Thus, the development and operation teams need to build a cooperative relationship. Culture, good practices, and tools for this are called DevOps [38]. First proposed around 2008, DevOps has become a common concept in software development. A similar framework, called MLOps, has been proposed for developing machine learning algorithms [39]. Quantum computing systems have concepts such as calibration that does not exist in traditional software development. Therefore, culture, good practices, and tools for quantum computing systems are necessary. Our research team calls them QCOps, including the following tools.

*1) OQTOPUS Admin*: OQTOPUS Admin is a GUI tool that manages users and devices. Administrators can view, suspend, or delete user information via the user management screen. They can also register, update, or delete information about quantum computers and simulators using the device management interface.

*2) QDash*: QDash is a web-based application that provides an intuitive interface for managing and monitoring calibration workflows, including multiple experiments. Although calibration is a critical process that determines the performance of the target quantum computer, its execution is inherently complex and labor-intensive owing to the time-varying nature of the qubit characteristics and intricate interdependencies among the quality-related parameters. Traditionally, calibration quality has depended on the expertise of individual researchers. To broaden access to quantum computation and reduce the operational burden, adopting a systematic approach that minimizes reliance on individual expertise is essential. QDash meets this requirement by offering a robust execution platform that collects, manages, and analyzes calibration data, improving reproducibility and efficiency in the calibration process.

Building a cloud-based quantum computing system requires either developing a large amount of a software independently or combining existing software while struggling with differences in specifications. OQTOPUS provides all necessary operation software and is expected to encourage new entrants and foster a quantum computing developer community. Furthermore, the key features implemented in the backend layer, which are close to a quantum computer, enable quantum computation to be performed effectively and efficiently.

## IV. USE CASES

This section presents the results of connecting OQTOPUS to a superconducting quantum computer already in operation at Osaka University using the following two jobs with QURI Parts OQTOPUS APIs available in [40].

Fig. 2. Sampling result displayed via OQTOPUS Frontend UI.

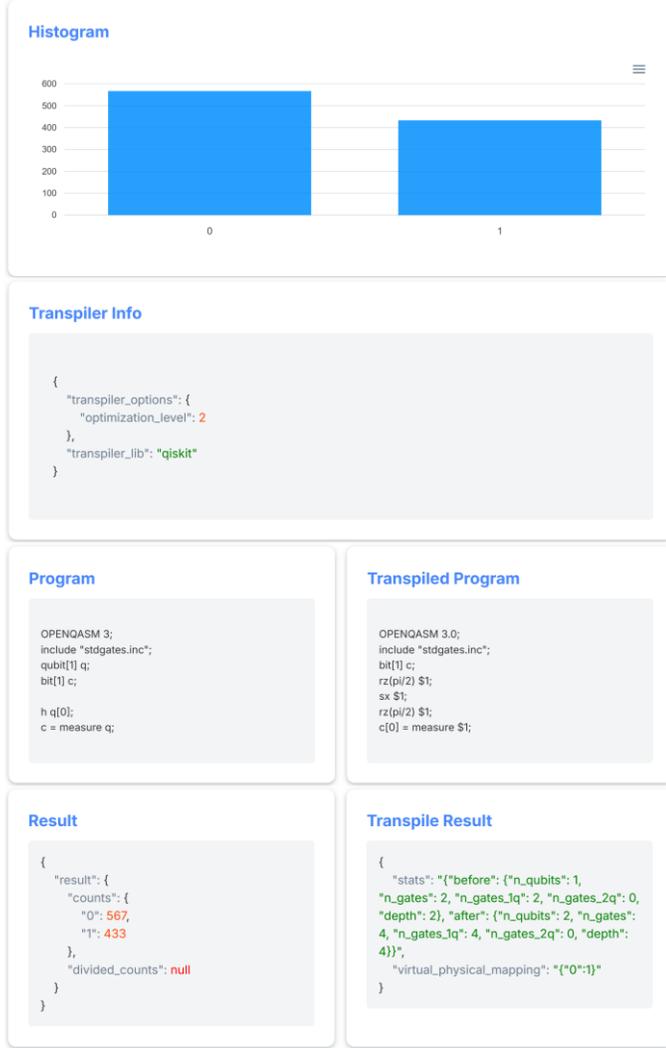

Fig. 3. Estimation result displayed using OQTOPUS Frontend UI.

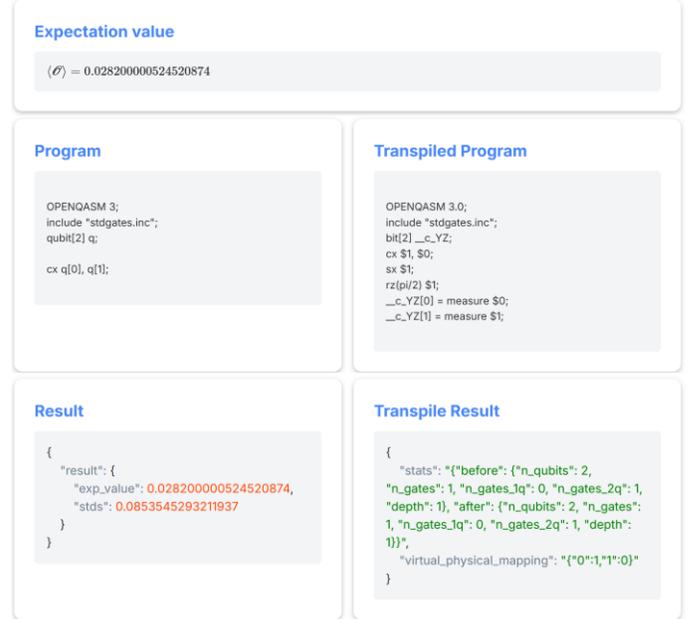

## A. Sampling Job

Listing 1 illustrates a user submitting a sampling job. In this example, a simple code written in QURI Parts OQTOPUS executes the sampling job on a quantum circuit inputted as "circuit" with 1000 shots specified in "shots" on our quantum computer identified as "device-id." The result of sampling running on a quantum computer via OQTOPUS is displayed to the user utilizing the web UI of OQTOPUS Frontend, as shown in Fig. 2. This figure confirms that the sampling result displayed in the histogram form is easy to understand. The codes after applying transpilation can also be viewed.

## B. Estimation Job

Listing 2 shows an example where a user submits an estimation job. In this example, a simple code written in QURI Parts OQTOPUS executes the estimation job of a quantum circuit and Pauli operators ($H = 1.5\,XX + 1.2\,YZ$) entered in "circuit" and "operator," respectively, with 1000 shots specified in "shots." The result of the estimation job running on a quantum computer via OQTOPUS is displayed to the user using the web UI of OQTOPUS Frontend, as shown in Fig. 3. Because the estimation result is an expected value, it is simply displayed.

These execution examples confirm that quantum computation can be performed from the user's environment through OQTOPUS by simply preparing hardware (a quantum chip, a pulse controller, etc.) and pulse sequencer software. Furthermore, users can easily understand the results of their experiments because OQTOPUS Frontend presents information clearly.

## V. CONCLUSION

Herein, we introduced the key features necessary to perform quantum computing effectively and attract users. Although it is desirable to implement these features in the backend layer for efficient computing, the backend layer is rarely made public. Therefore, we developed and released a sophisticated cloud-based quantum computing system that included the backend layer with all these key features. Furthermore, our system was equipped with all basic operation software required to provide it as a cloud service, making it easy to use for institutions and companies seeking to enter the quantum computing field. We expect OQTOPUS to contribute to the expansion of this field.

In our future works, we want to make all system software related to quantum computing publicly available. Our priority is to develop a quantum control system and system monitoring tools as open-source software.

ACKNOWLEDGMENT

This work was supported by JST COI-NEXT, Grant No.JPMJPF2014.